\documentclass{pasj00}

\SetRunningHead{S. Konami et al.}{
Abundance patterns in the ISM of NGC~1316}

\title{Abundance Patterns in the Interstellar Medium of the S0 Galaxy 
NGC~1316 (Fornax A) Revealed with Suzaku}
\author{
 Saori \textsc{Konami},\altaffilmark{1,2}
 Kyoko \textsc{Matsushita},\altaffilmark{1} 
 Ryo \textsc{Nagino},\altaffilmark{1} 
Makoto S. \textsc{Tashiro},\altaffilmark{3}
Toru \textsc{Tamagawa},\altaffilmark{2,1}\\
and Kazuo \textsc{Makishima},\altaffilmark{4,2}
}
\altaffiltext{1}{
Department of Physics, Tokyo University of Science,
1-3 Kagurazaka, Shinjuku-ku, Tokyo 162-8601}
\altaffiltext{2}{
Cosmic Radiation Laboratory, the Institute of Physical and Chemical Research,
2-1 Hirosawa, Wako, Saitama 351-0198}
\email{konami@crab.riken.jp}
\altaffiltext{3}{
Department of Physics, Saitama University,
225 Shimo-Okubo, Sakura, Saitama 338-8570}
\altaffiltext{4}{
Department of Physics, University of Tokyo,
7-3-1, Hongo, Bunkyo-ku, Tokyo 113-0033}

\Received{}
\Accepted{}
\Published{---}

\begin{document}
\maketitle
\begin{abstract}
The Suzaku X-ray satellite observed the nearby S0 galaxy NGC~1316, 
a merger remnant aged 3 Gyr. 
The total good exposure time was 48.7 ksec. 
The spectra  were
well represented by a two-temperature thermal model for the 
interstellar medium (ISM) plus a power-law model. 
The cool and hot temperatures of the thermal model were 0.48$\pm{0.03}$ and 
0.92$\pm0.04$ keV, respectively. The excellent spectral sensitivity of Suzaku 
enables for the first time to measure the metal abundances of O, Ne, Mg, Si, 
and Fe in the ISM. 
The resultant abundance pattern of O, Ne, Mg, Si, and Fe 
is close to that of  the new solar abundance determined by \citet{lodders_03}.
The measured abundance pattern is compared with those of
elliptical galaxies and an S0 galaxy, observed with Suzaku.
Considering the metal-enrichment from present Type Ia supernovae,
the near-solar abundance pattern of the ISM in NGC~1316 indicates an enhanced 
$\alpha/{\rm Fe}$  ratio of stellar materials in the entire galaxy, like in 
giant elliptical galaxies.
\end{abstract}

\section{Introduction}
 
Hot X-ray emitting interstellar medium (ISM) carries important information
about the history of star formation and evolution of galaxies. 
The metals in the ISM in early-type galaxies have been enriched by type Ia 
supernovae (SNe Ia)  and stellar mass loss. 
Since SNe Ia do not produce O or Mg, abundances of these elements in the ISM 
reflect stellar abundances. 
Thus, the abundance pattern of the ISM can play a key role in studying the origin 
of metals.

The ASCA satellite first enabled us to measure the metal abundances 
in the ISM of the early-type galaxies (e.g., 
\cite{awaki_94}; \cite{loewenstein_94}; \cite{mushotzky_94}; 
\cite{matsushita_94}). 
These galaxies have about 1 solar iron abundance, employing
a multi-temperature plasma model \citep{buote_98}, or
considering uncertainties in the  Fe-L atomic
data (\cite{arimoto_97}; \cite{matsushita_97}; \cite{matsushita_00}).
Chandra and XMM-Newton measured the metal abundances and spatial 
distributions of metals in the ISM of the early-type galaxies, because of 
their large effective area and good angular resolution 
(\cite{humphrey_04}; \cite{humphrey_06}; \cite{werner_06}; \cite{tozuka_08}). 
In particular NGC~5044, NGC~1399, and NGC~4636, the abundances of non-Fe
elements have good constraints with Chandra and XMM (\cite{buote_03}; \cite{ji_09}). 
The observed abundance patterns are consistent with the solar ratio. 
However, O and Mg abundances 
of relative faint X-ray galaxies
still remain highly uncertain, 
due to a highly asymmetric energy response in the low-energy 
region below 1 keV and a strong instrumental Al line of these missions, 
respectively.  
The Suzaku XIS \citep{koyama_07} has an improved line spread function due to 
a very small low-pulse-height tail below 1 keV, with a lower and more stable 
background level compared to the Chandra ACIS and the XMM-Newton EPIC.
The abundance patterns in the ISM of several giant elliptical galaxies, 
including NGC~1399 and NGC~1404 \citep{matsushita_07}, NGC~720 
\citep{tawara_08}, 
and NGC~4636 \citep{hayashi_09}, have
been revealed by recent Suzaku observations.
The derived Fe abundances of the ISM
are about 1 solar, and the abundance patterns 
are close to that of the  new solar abundance determined by \citet{lodders_03}.
From these results, the contribution 
of SNe Ia and stellar mass loss to the 
metal enrichment of the ISM and intracluster medium  were discussed.
However, NGC~4382 is the only S0 galaxy whose ISM abundance pattern  
has so far been derived with Suzaku \citep{nagino2009}.
The O/Fe ratio of the ISM in this
galaxy is smaller by a factor of two as compared to
than those in the four elliptical galaxies.

In terms of the cosmology dominated by cold dark matter,
it is considered that at least some elliptical galaxies 
were formed by a hierarchical formation scenario, in which larger 
spheroidals were assembled relatively late through mergers of the 
late-type galaxies of comparable
mass  (\cite{delucia_06} and references therein).
In contrast, S0 galaxies in clusters of galaxies are considered to be formed through 
 morphological transformations, from  infalling gas-rich spiral galaxies 
to gas-poor S0 galaxies. This is because spiral galaxies were much more common, 
and S0 galaxies were much
rarer, in distant than nearby galaxy clusters 
(e.g. \cite{poggianti_09}; \cite{desai_07}; \cite{smith_05}; 
\cite{postman_05}).
In spite of these differences, 
the elliptical and S0 galaxies have been treated together in most of the past studies.
However, since there are possible differences in the formation scenario, 
we need to study the elliptical and S0 galaxies separately.

The abundance pattern of stars reflects their formation history,
since a longer formation time provides a higher concentration
of trapped SNe Ia products to the stars.
If the elliptical and S0 galaxies have different formation histories, 
their stellar metallicity and/or present SN Ia rate may be different.
The stellar metallicity of the early-type galaxies
 has often been investigated in optical observations
of their central regions  
(e.g., \cite{thomas_05}; \cite{bedregal_08}; \cite{walcher_09}).
However, there may be systematic uncertainties in the assumption of the
age-population of stars, and in atomic physics.
Also the values of solar abundances have changed enormously, 
considering the three-dimensional hydrostatic atmosphere and 
non-local thermodynamic equilibrium \citep{asplund_05}.
In contrast, X-ray observations of the ISM, show much simpler atomic data and 
temperature structures as compared to those in the optical spectra, and
we can observe the entire region of each galaxy,  
although the metals in the ISM are a mixture of those
from stars and recent SNe Ia. Considering these, further X-ray studies of 
S0 galaxies are thought to be important.

NGC~1316 is a peculiar S0 galaxy, with  numerous tidal tails. This galaxy has 
a surprisingly high central surface brightness, and a 
lower velocity dispersion ($\sigma$ = 227$\pm$33 km s $^{-1}$; 
\cite{goud_01a})  for its galaxy luminosity 
than other elliptical galaxies with similar luminosities 
(e.g., \cite{schweizer_80}; \cite{schweizer_81}; \cite{donofrio_97}).
Due to these features, NGC~1316 has been thought to have undergone a 
major merger relatively recently. The merger age is estimated to be 
$\sim$3 Gyr, based on the age of the bright globular clusters 
 \citep{goud_01a}. 
Therefore, NGC~1316 is good target for investigation of the galaxy 
formation process through mergers. 
Measurements of the absorption-line indices of the central region, or 1/10 $r_e$
of  NGC~1316, indicates a small overabundance of $\alpha$-to-Fe ratio
of [$\alpha/{\rm Fe}$]=0.15 (\cite{thomas_05}), which is
smaller than the [$\alpha/{\rm Fe}$]$\sim$0.3, in center of the 
giant elliptical galaxies (e.g. \cite{thomas_05}; \cite{pipino_09}).
Here, $r_e$ is the effective radius of the galaxy.
These line indices also indicate a somewhat younger age of 3--5 Gyr
\citep{bedregal_08}.

ROSAT observations of NGC~1316 detected a thermal emission component from 
hot ISM \citep{kim_98}, 
using Chandra data of an good exposure of 25 ksec, \citet{kim_03} found 
that the ISM has a temperature in the 0.5--0.6 keV range and a 0.3--8 keV 
luminosity of 3.1$\times 10^{40}$ erg s$^{-1}$. 
The metal abundance of the ISM were reported to be 0.2--1.3 solar adopting 
the solar abundance table by \citet{angre_89}.
NGC~1316 hosts radio lobes, that generate inverse-Compton X-rays 
as well as synchrotron radio emission 
(e.g. \cite{feigelson_95}; \cite{kaneda_95}; \cite{isobe_06}).
Although this means that NGC~1316 harbors an active galactic nucleus (AGN), 
\citet{iyomoto_98} and \citet{kim_03} showed that the present activity 
of the AGN is very low.
A recent Suzaku observation detected the inverse-Compton X-ray up to 
20 keV utilizing Suzaku from the west lobe \citep{tashiro_09}.

In this paper, we analyze Suzaku XIS data of NGC~1316. We use a distance of 18.6 Mpc 
to NGC~1316 based on the Hubble Space Telescope measurements of Cepheid variables 
\citep{madore_99}, and a 
redshift of  0.005871 including proper motion \citep{longhetti_98}.
We use the new solar abundances in \citet{lodders_03}. 
Unless noted otherwise, the quoted errors are refer to 90\% confidence interval 
for a single interesting parameter. 

\begin{figure}
\begin{center}
\begin{minipage}{0.5\textwidth}
\centerline{
\FigureFile(\textwidth,\textwidth){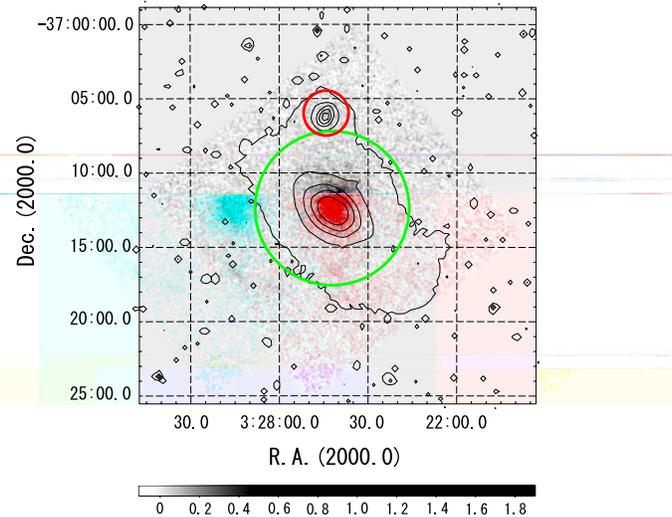}}
\label{figure1}
\caption{
The 0.4--5.0 keV XIS image of NGC~1316, overlaid on 
an optical contour map taken by Digitized Sky Survey. 
The observed XIS0, 1, and 3 images 
were added on the sky coordinate, and smoothed with a $\sigma$ = 5.2 arcsec 
Gaussian profile. Estimated contributions of the cosmic X-ray background 
(CXB) and the instrumental non-X-ray background (NXB) were subtracted, 
although vignetting was not corrected. The region used for spectrum 
extraction is indicated by a green circle. NGC~1317, expressed by 
a reg circle, was excluded from the background and galaxy spectra.
}
\end{minipage}
\end{center}
\end{figure}

\begin{table*}
\caption{
Best-fit parameters for the apec components + power-law models.$^{\ast}$}
\label{table1}
\begin{center}
\begin{tabular}{lcccc} \hline\hline
Parameters               &             & (i)                       & (ii)   \\ \hline
$\Gamma_{\rm OTHERS}$        &             & 1.6 (fix)                 & 1.6 (fix)  \\ 
$Norm_{\rm OTHERS}\,^\dagger$ &             & 1.10$\pm{0.07}$      & 1.10$\pm{0.07}$  \\ 
$N_{\rm H}$               & (cm$^{-2}$) & 1.92$\times 10^{20}$ (fix)  & 1.92$\times 10^{20}$ (fix) \\
 [1.0ex]
$kT_{\rm ICM}$            & (keV)       & 0.74$^{+0.06}_{-0.05}$        & 0.74$^{+0.06}_{-0.05}$  \\ 
Abundance                & (solar)     & 0.3 (fix)                     & 0.3 (fix) \\ 
$Norm_{\rm ICM}\,^\ddagger$ &             & 1.88$\pm{0.36}$        & 1.91$\pm{0.35}$ \\
[1.0ex] 
$kT_{\rm MWH}$            & (keV)       & 0.30$^{+0.03}_{-0.04}$        & 0.30$^{+0.03}_{-0.04}$  \\ 
Abundance                & (solar)     & 1 (fix)                   & 1 (fix) \\ 
$Norm_{\rm MWH}\,^\ddagger$ &             & 0.77$\pm{0.17}$        & 0.68$\pm{0.15}$ \\
 [1.0ex]
$kT_{\rm LHB}$             & (keV)       & 0.1 (fix)                   & 0.1 (fix) \\ 
Abundance                 & (solar)     & 1 (fix)                     & 1 (fix) \\ 
$Norm_{\rm LHB}\,^\ddagger$  &             & 2.79$^{+0.59}_{-0.58}$        & 2.73$\pm{0.60}$ \\
[2.0ex] 
${\chi^2}$/d.o.f.         &             & 490/359                     & 490/359 \\ \hline \hline
\end{tabular}
\end{center}
\parbox{\textwidth}{\footnotesize
\footnotemark[$\ast$]
The apec components for spectra in the background region of 
NGC~1316 with the ICM component of the Fornax cluster, absorbed or non-absorbed 
MWH component, LHB component for the Galactic emission, and 
a power-law model for CXB. \\
\footnotemark[$\dagger$] 
Normalization of the power-law component
divided by the solid angle, $\Omega^{\makebox{\tiny\sc u}}$,
assumed in the uniform-sky ARF calculation (20$'$ radius),
in units of 10$^{-3}$ $\Omega^{\makebox{\tiny\sc u}}$ photons 
keV$^{-1}$ cm$^{-2}$ s$^{-1}$ arcmin$^{-2}$ at 1 keV.\\
\footnotemark[$\ddagger$]
Normalization of the apec components
divided by the solid angle same as the normalization of apec,
${\it Norm} = \int n_{\rm e} n_{\rm H} dV \,/\,
(4\pi\, (1+z)^2 D_{\rm A}^{\,2}) \,/\, \Omega^{\makebox{\tiny\sc u}}$
$\times 10^{-17}$ cm$^{-5}$~arcmin$^{-2}$, 
where $D_{\rm A}$ is the angular distance to the source.}
\end{table*}

\section{Observation and Data Reduction}\label{sec:obs}
Suzaku observed NGC~1316 in 2006 December, using the XIS \citep{koyama_07}.
The XIS consists of two 
front-illuminated (FI: XIS0 and XIS3) CCD cameras and one back-illuminated 
(BI: XIS1) CCD camera. 
The averaged pointing direction of the XIS was at 
($\alpha$, $\delta$)=(\timeform{3h22m40.4s}, \timeform{-37D12'10.4''})\@, 
right on the NGC~1316 nucleus. 
The XIS was operated in normal 
clocking mode (8~s exposure per frame), with the standard 
5 $\times$ 5 and 3 $\times$ 3 editing mode. 
We processed the XIS data using the ``xispi'' and ``makepi'' ftool tasks 
and CALDB files of 2008-08-25 version. 
Then, the XIS data were cleaned by assuming thresholds on the Earth 
elevation angle of $>~5^{\circ}$ and the Day-Earth elevation angle of 
$>~20^{\circ}$. 
We also discarded data taken when the time after a spacecraft exit from 
the south Atlantic anomaly is less than 436 sec. 
We created a 0.5--2 keV light curve for each sensor, with 540 sec binning, 
to exclude periods of anomalous event rates higher or less 
than $\pm 3\sigma$ from the mean. 
After this screening, the remaining good exposures were 48.7 ksec 
for both FIs and BI.
Event screening with cut-off rigidity was not performed.

The spectral analysis was performed with HEAsoft version 6.5 and XSPEC 12.4.
To subtract the non-X-ray background (NXB), we employed 
the Dark-Earth database using the ``xisnxbgen'' ftool task \citep{tawa_08}.
We generated two different ancillary response files (ARFs) 
for the spectrum of each region; one assumed uniform 
sky emission, while the other utilized the observed XIS1 image by the 
``xissimarfgen'' ftool task \citep{ishisaki_07}. In the ARFs, we also 
included the effect of contamination on the optical blocking filters of the XIS.

\begin{figure*}
\begin{center}
\centerline{
\FigureFile(\textwidth,\textwidth){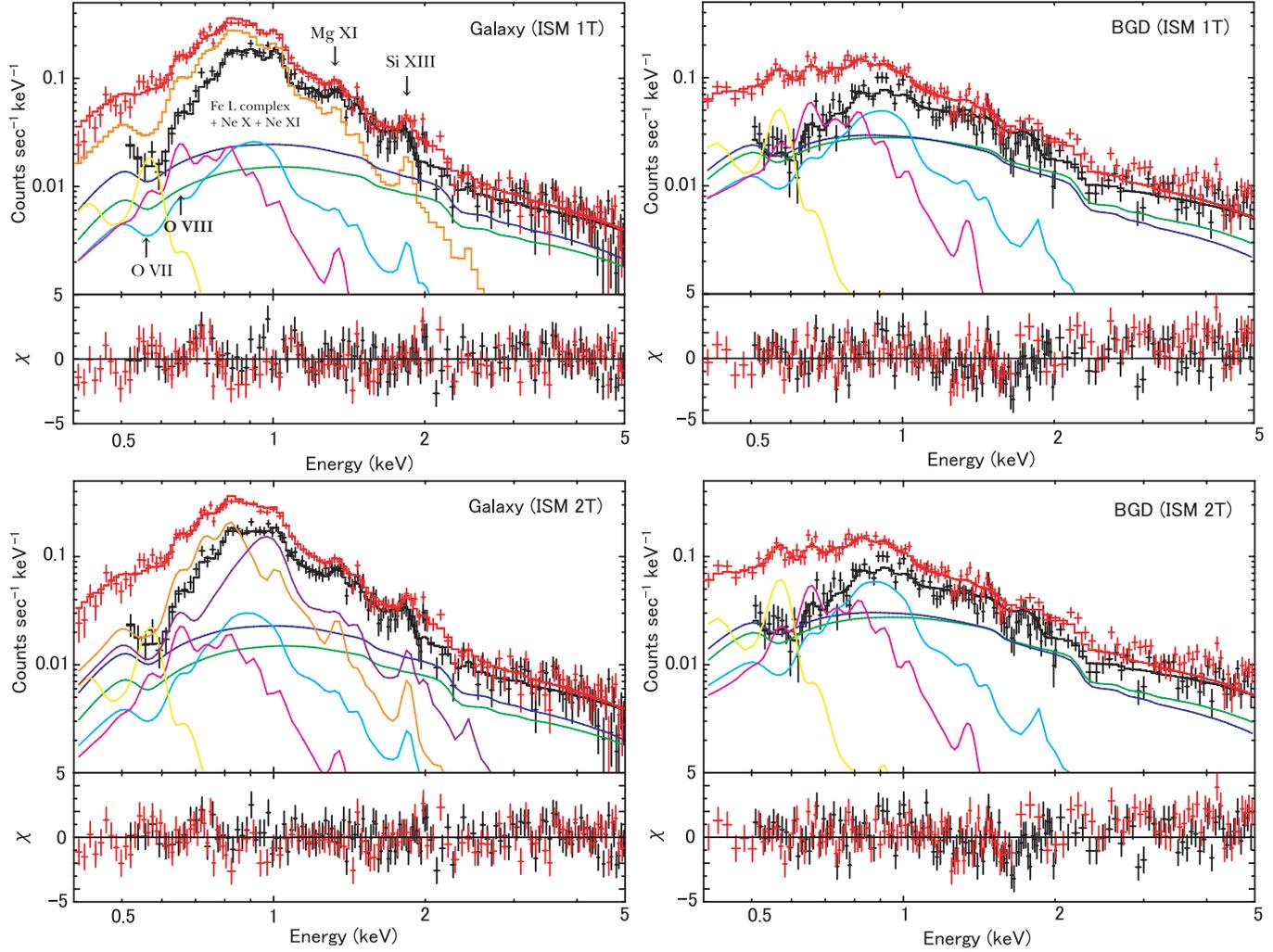}}
\caption{
NXB-subtracted XIS0 (black) and XIS1 (red) spectra 
of NGC~1316 (left column), and those of the background region (right column), 
shown without removing the instrumental responses. 
The top two and bottom two panels employ the 
one- and two-temperature model for the ISM, respectively.
Black and red lines show the best-fit model for the XIS0 and XIS1, 
respectively. For simplicity, only the model components for XIS1 spectra 
are shown. Orange and purple lines are the ISM components, 
cyan is emission from the ICM of Fornax cluster by apec$_{\rm ICM}$,
magenta and yellow are the Galactic background emission by apec$_{\rm MWH}$ 
and apec$_{\rm LHB}$, while blue and green are the CXB and (LMXB + radio lobe) 
components, respectively. 
The background components, except power-law$_{\rm OTHERS}$, are common between 
the on-source and background spectra, but scaled to the respective 
data accumulation area.
}\label{figure2}
\end{center}
\end{figure*}

\section{Analysis and Results}
\subsection{Extraction Regions of Spectra}
\label{x-ray}

The observed X-ray image in the 0.4--5 keV range is shown in figure 1, 
superposed on an optical contours from the Digitized Sky Survey. 
We used the following two regions to extract spectra; 
one is NGC~1316 region, which is a circular region with a 5$'$ radius shown 
in a green circle in figure 1. 
The other is background region, which is the entire XIS field of 
view excluding the NGC~1316 region. 
We excluded the region around  NGC~1317, shown in a red circle
in figure 1.
In addition, three calibration sources, with emission peaks 
at 5.9 keV, are located at the corners of the XIS. 
We included these regions to improve the photon statistics 
because the ISM  emits almost no photons above 5 keV.

\begin{table*}
\caption{
Summary of the best-fit parameters for the NGC~1316 and the background region with 1T for ISM model, 
2T for ISM model, and 2T for ISM adding 10\% systematic error of Fe-L model.
}
\label{table2}
\begin{center}
\begin{tabular}{lccccc} \hline \hline
Parameters && 1T for ISM & 2T for ISM &  Fe-L \\ \hline
$kT_{\rm 1T}$ & (keV) & 0.62$\pm{0.01}$ & 0.48$\pm{0.03}$ &  0.40$^{+0.08}_{-0.03}$ \\ 
$Norm_{\rm 1T}\,^\ast$ & & 2.10$^{+0.17}_{-0.18}$ & 1.04$^{+0.21}_{-0.24}$ &  0.95$^{+0.16}_{-0.26}$ \\ 
$kT_{\rm 2T}$ & (keV) &-& 0.92$\pm{0.04}$ &  0.93$^{+0.05}_{-0.04}$ \\ 
$Norm_{\rm 2T}\,^\ast$ & &-& 0.80$^{+0.16}_{-0.19}$ &  0.70$^{+0.12}_{-0.22}$ \\ 
O & (solar) & 0.71$^{+0.41}_{-0.23}$ & 0.79$^{+0.51}_{-0.31}$ &  0.71$^{+0.50}_{-0.26}$ \\
Ne & (solar) & 1.31$^{+0.57}_{-0.35}$ & 0.63$^{+0.53}_{-0.45}$ &  0.66$^{+0.75}_{-0.48}$\\
Mg, Al & (solar) & 0.43$^{+0.22}_{-0.13}$ & 0.55$^{+0.35}_{-0.23}$ &  0.77$^{+0.81}_{-0.33}$\\
Si,S,Ar,Ca & (solar) & 0.31$^{+0.20}_{-0.12}$ & 0.48$^{+0.32}_{-0.38}$ &  0.64$^{+0.66}_{-0.25}$ \\
Fe, Ni & (solar) & 0.44$^{+0.16}_{-0.09}$  & 0.73$^{+0.42}_{-0.15}$ &  1.01$^{+0.75}_{-0.31}$ \\
O/Fe & (solar) & 1.61$^{+0.43}_{-0.47}$ & 1.08$^{+0.28}_{-0.41}$ &  0.70$^{+0.51}_{-0.25}$ \\
Ne/Fe & (solar) & 3.00$\pm{0.57}$ & 0.86$^{+0.68}_{-0.52}$ &  0.65$^{+0.83}_{-0.52}$\\
Mg/Fe & (solar) & 0.98$^{+0.23}_{-0.27}$  & 0.75$^{+0.22}_{-0.25}$ & 0.76$^{+0.26}_{-0.06}$ \\
Si/Fe & (solar) & 0.70$^{+0.30}_{-0.27}$ & 0.66$^{+0.21}_{-0.23}$ & 0.63$^{+0.31}_{-0.20}$\\[2.0ex]
$kT_{\rm ICM}$ & (keV) & 0.77$^{+0.08}_{-0.06}$ & 0.72$^{+0.05}_{-0.04}$  & 0.72$^{+0.04}_{-0.06}$ \\ 
$Norm_{\rm ICM}\,^{\ast}$ & & 2.51$\pm{0.51}$ & 2.83$\pm{0.36}$  & 2.84$^{+0.44}_{-0.36}$ \\ 
$kT_{\rm MWH}$ & (keV) & 0.30$\pm{0.03}$  & 0.28$^{+0.02}_{-0.03}$ & 0.28$\pm{0.02}$  \\ 
$Norm_{\rm MWH}\,^{\ast}$ & & 1.44$^{+0.27}_{-0.26}$ & 1.26$^{+0.23}_{-0.28}$ & 1.23$^{+0.21}_{-0.14}$  \\ 
$kT_{\rm LHB}$ & (keV) & 0.1 (fix) & 0.1 (fix) & 0.1 (fix)  \\ 
$Norm_{\rm LHB}\,^{\ast}$ & & 4.52$\pm{0.98}$ & 4.50$^{+1.00}_{-1.40}$ & 4.37$^{+1.41}_{-0.98}$  \\[2.0ex]
${\chi^2}$/d.o.f. & & 970/703 & 921/701 & 859/692 \\ \hline \hline
\end{tabular}
\end{center}
\parbox{\textwidth}{\footnotesize
\footnotemark[$\ast$] 
Normalization of the vapec and apec components scaled with a factor 
of {\sc source\_ratio\_reg} / {\sc area}, 
which is $Norm=\frac{\makebox{\sc source\_ratio\_reg}}
{\makebox{\sc area}} \int n_{\rm e} n_{\rm H} dV 
\,/\, [4\pi\, (1+z)^2 D_{\rm A}^{\,2}]$ $\times 10^{-17}$~cm$^{-5}$
~arcmin$^{-2}$, where $D_{\rm A}$ is the 
angular distance to the source.\\
} 
\end{table*}

\subsection{Estimation of Background Spectra}
\label{back}

In order to estimate the background components, we first fitted the spectra 
extracted over the  background region. 
Since the NXB component was already subtracted, the remaining background consisted 
of extra-galactic cosmic X-ray background (CXB), and emission 
from our Galaxy.
We assumed a power-law model for the CXB component, and a two 
temperature model from the  Galactic emission.
Empirically, 
one thermal component represents the sum of solar wind charge exchange
(SWCX) and local hot bubble (LHB),
while Milky Way halo (MWH) emits the other thermal component \citep{yoshino_09}.
We also added a thin thermal plasma emission from intracluster medium (ICM)
of the Fornax cluster \citep{tashiro_09}. 
We thus, fitted the spectra with the following model :
phabs $\times$ (power-law + apec$_{\rm ICM}$ + apec$_{\rm MWH}$) + 
apec$_{\rm LHB}$.
Here, the ``phabs'' factor represents the photoelectric-absorption,
whose  column density was  fixed to the Galactic value of 1.92$\times10^{20}$ cm$^{-2}$
in the direction of NGC~1316. 
The ``apec$_{\rm ICM}$'', ``apec$_{\rm MWH}$'' and ``apec$_{\rm LHB}$'' terms 
mean the  thermal emission, in terms of APEC 
plasma model \citep{smith_01}, from the ICM, MWH, and LHB (and SWCX), 
respectively.
We assumed that the apec$_{\rm ICM}$ has a metal abundance fixed at 0.3 solar, 
while apec$_{\rm MWH}$ and apec$_{\rm LHB}$ 1.0 solar. 
The redshift of apec$_{\rm ICM}$ were also fixed at 0.005871,
while apec$_{\rm MWH}$ and apec$_{\rm LHB}$ zero.
The power-law slope of the CXB emission was fixed at $\Gamma=1.4$ 
\citep{kushino_02}.
The spectra from the BI and FI CCDs were simultaneously fitted in 
the 0.4--5.0 and 0.5--5.0 keV range, respectively. 

This model well reproduced the spectra, with ${\chi^2}$/d.o.f. $=$ 455/360. 
However, the strength of the power-law representing the CXB was about twice 
as high as that of \citet{kushino_02}.
We consider that this is due to contributions by other sources in NGC~1316,
such as the radio lobes and integrated low mass X-ray binaries (LMXBs). 
The index of the power-law component from  the radio-lobes 
is 1.68 \citep{tashiro_09}.
The integrated spectra from 
discrete sources in the early type galaxies are approximated with a 
power-law model with an index of 1.6
(e.g., \cite{xu_05}, \cite{randall_06}).
Therefore, we added a power-law component, 
``power-law$_{\rm OTHERS}$'', with its index fixed at 1.6, and then fitted
the spectra  with the following model (i):
phabs $\times$ (power-law$_{\rm CXB}$ + power-law$_{\rm OTHERS}$ + apec$_{\rm ICM}$ 
+ apec$_{\rm MWH}$) + apec$_{\rm LHB}$.
The index and normalization of power-law$_{\rm CXB}$ were fixed 
at the values by \citet{kushino_02}. 
The results of the fit are summarized in table \ref{table1}.

We also examined the following model (ii):
phabs $\times$ (power-law$_{\rm CXB}$ + power-law$_{\rm OTHERS}$ + 
apec$_{\rm ICM}$) + apec$_{\rm MWH}$ + apec$_{\rm LHB}$.
The difference of this model from model (i) is whether or not the MWH component is
absorbed by the Galactic column 
(e.g., \cite{yamasaki_09}; \cite{konami_09}). 
We summarize the results in table \ref{table1}.
There was no significant difference in the derived parameters
between model (i) and (ii). 
Hereafter, we adopted model (i) as the background model 
for the spectral fitting, unless otherwise stated.

To evaluate the Galactic 
emission, we also fitted the O\emissiontype{VII} and O\emissiontype{VIII} 
lines of the spectra of the background region with Gaussians and 
a power-law model in the 0.5--0.7 keV range.
The resultant intensities of the O\emissiontype{VII} and
O\emissiontype{VIII}, without correcting for absorption, 
are 9.2$^{+5.7}_{-4.3}$ and 2.4$^{+0.7}_{-0.8}$ photons 
cm$^{-2}$ s$^{-1}$ sr$^{-1}$, 
respectively. These values are consistent with the previously reported values 
from other sky region
\citep{mccammon_02, sato_07, yamasaki_09, konami_09}.  
Thus, the Galactic foreground emission in the present sky direction is 
considered typical.

\begin{figure*}
\begin{center}
\centerline{
\FigureFile(\textwidth,\textwidth){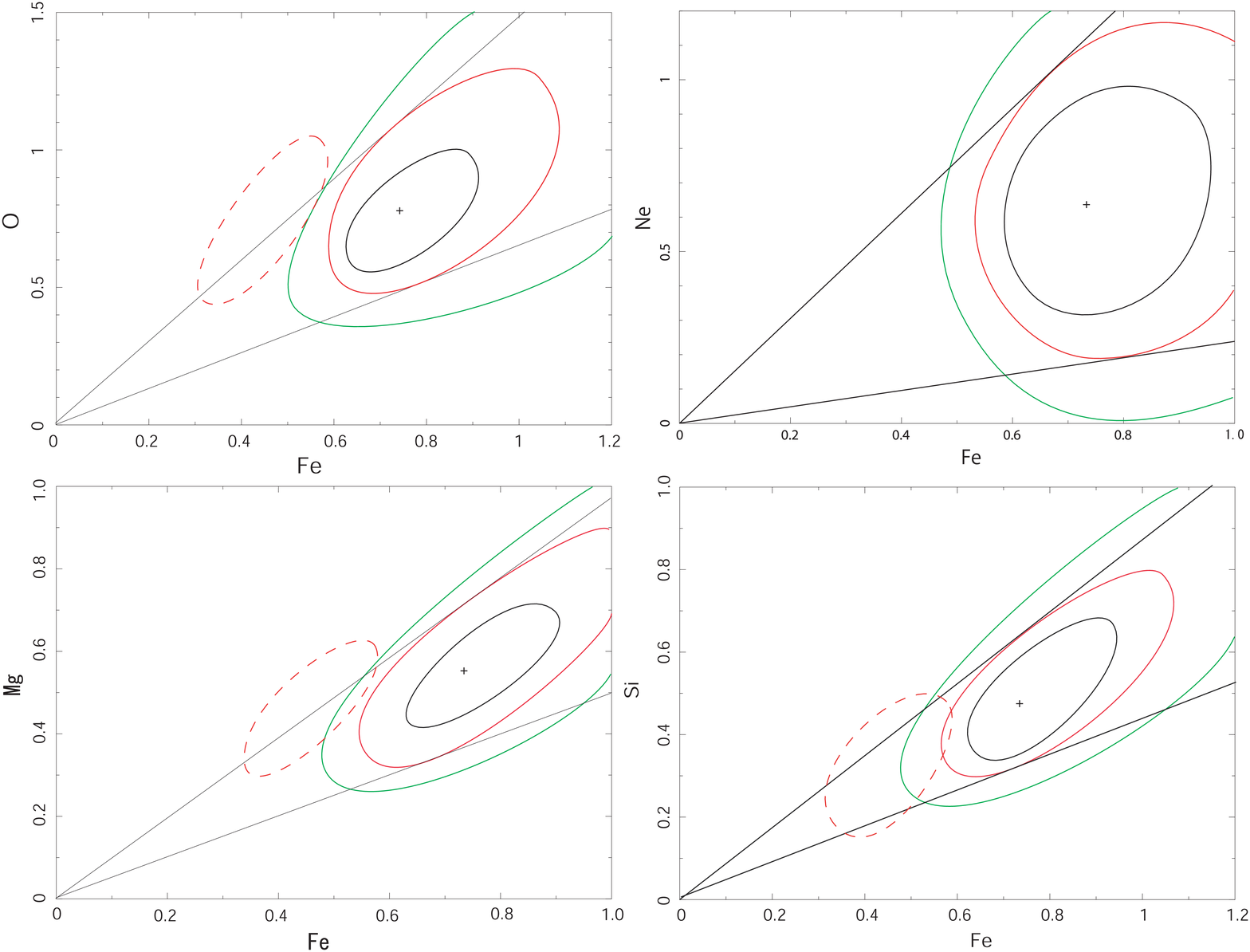}}
\caption{
Confidence contours between the metal (O, Ne, Mg, and Si) and Fe 
abundances, determined with the two-temperature model for the ISM.
The black, red, and green contours represent 68\%, 90\%, and 99\% confidence ranges, 
respectively. The red dashed contours indicate 
90\% confidence range determined with the one-temperature model  for  the ISM. 
}\label{figure3}
\end{center}
\end{figure*}

\subsection{Emission from  NGC~1316}
\label{gal}

We fitted the spectra of the NGC~1316 region with the model: 
phabs$\times$vapec$_{\rm ISM}$ + background models.
In this model, ``phabs'' represents the Galactic 
absorption in the direction of NGC~1316, fixed at 
$N_{\rm H} = 1.92\times10^{20}$ cm$^{-2}$. 

The ISM emission of NGC~1316, vapec$_{\rm ISM}$, was represented by 
one-temperature (1T) or two-temperature (2T) models,
employing the vapec code \citep{smith_01}. 
The abundances of He, C, and N were fixed 
to the solar value. We also divided the other metals into 
five groups as O, Ne, (Mg \& Al), (Si, S, Ar, Ca), and (Fe \& Ni), 
based on the metal synthesis mechanism of SNe, and allowed 
each group to vary.
In the 2T modeling, the two vapec components were assumed to share the 
same elemental abundances.
Using this model, 
we simultaneously fitted the spectra of the  NGC~1316 and background regions
using  the energy ranges of 0.4--5.0 keV for the BI CCD and 
0.5--5.0 keV for the FI CCDs.
The normalization of power-law$_{\rm OTHERS}$ was not linked between  the
two regions, because the number of LMXBs may  
increase toward the galactic center, and the brightness of the radio-lobes may 
be different.
All the other background components were constrained to be the same 
between the two regions.
The normalizations of the ISM components in the background region 
were fixed at 0.

The results of these fits are summarized in figure \ref{figure2} and 
table \ref{table2}.
Several emission lines seen around 0.5--0.6 keV, 0.6--0.7 keV, 
$\sim$ 1.3 keV, and $\sim$ 1.8 keV are identified with K$\alpha$ lines of 
O \emissiontype{VII}, O \emissiontype{VIII}, 
Mg \emissiontype{XI}, and Si \emissiontype{XIII}, respectively. 
The emission bump around 0.7--1 keV corresponds to 
Fe-L complex, as well as to K-lines from Ne \emissiontype{IX} and 
Ne \emissiontype{X}. 
These fits are not formally acceptable, mainly,  because of discrepancies between 
the FI and BI spectra of the background region above 2 keV.
Although this discrepancy might be related to some remaining problems 
with the NXB subtraction, they are unlikely to affected the ISM modeling.
The 1T model results in  $kT=0.62$ keV\@. 
The fit statistics shown in table \ref{table2} 
clearly favor the 2T model, which 
gives the two temperatures as 0.48 and 0.92 keV\@. 
The abundance of Fe does not depend very much on the
temperature structure of the ISM.
Adoption of the 2T model shows, the Fe abundance of the ISM as 0.58--1.15 solar.
This value is consistent with the ISM metallicity of 0.4--2.1 solar 
derived from the Chandra observation derived by \citet{kim_03}. 
Here, the value was converted using the solar abundance table by \citet{lodders_03}.

\begin{figure*}
\begin{center}
\centerline{
\FigureFile(\textwidth,\textwidth){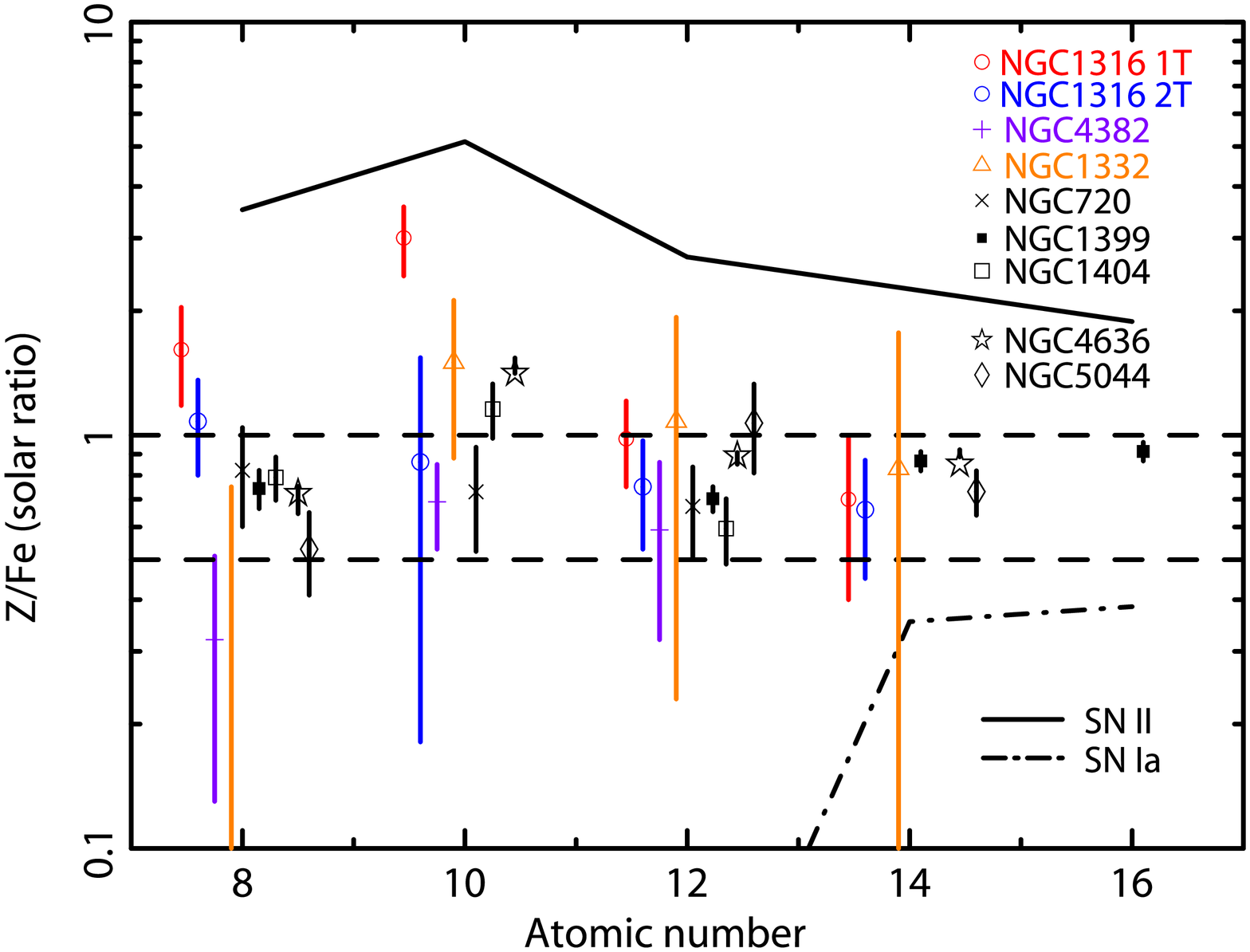}}
\caption{
Abundance ratios of O, Ne, Mg, and Si to Fe for the one-temperature (red) and 
two-temperature (blue) model of the ISM in NGC~1316. 
Abundance patterns of NGC~4382 (purple; \cite{nagino2009}), 
NGC~1332 (orange; \cite{humphrey_06}),
NGC~720 (black crosses; \cite{tawara_08}), NGC~1399 (black closed squares; \cite{matsushita_07}), 
NGC~1404 (black open squares; \cite{matsushita_07}), NGC~4636 
(black stars; \cite{hayashi_09}), and NGC~5044 (black diamond; 
\cite{buote_03} and \cite{komiyama_09}) are also shown. 
Solid and dot-dashed lines represent the number ratios of metals to 
Fe for the SN II and SN Ia products \citep{iwamoto_99, nomoto_06}.
}\label{figure4}
\end{center}
\end{figure*}

In order to examine the abundance ratios rather than their absolute values, 
we calculated confidence contours between the abundances of 
metals (O, Ne, Mg, and Si) and of Fe, 
using the 2T model for the ISM. 
The results are shown in figure \ref{figure3} and table \ref{table2}.
The elongated shape of the confidence contours indicates that the relative 
values are determined more accurately than the absolute values.
Figure \ref{figure4} the shows metal-to-Fe ratios in NGC~1316 using the 
1T and 2T models for the ISM, which were derived from the two-parameter 
confidence contours shown in figure \ref{figure3}. 
The O/Fe ratio varies  from 1.7 with the 1T model to 1.1 with the 2T model
in solar units. 
On the other hand, the Ne/Fe ratio changes considerably, from 
$\sim$3 solar with the 1T model to $\sim$1 solar with the 2T case.
We note that the Ne abundance is not reliably determined 
due to an overlap with the strong and complex Fe-L line emissions.
The 1T and 2T models give similar values for the
Mg/Fe and Si/Fe ratios.

\subsection{Systematic Uncertainties in the Abundance Ratios}

We also tried to fit the same spectra with the three-temperature vapec model for the ISM.
The derived temperature of the third component has a large error, 
and the Fe abundance increased to 1.03$^{+0.53}_{-0.29}$ in solar units.
Meanwhile, the 
derived abundance ratios are consistent with those from the 2T model.
To examine the dependence on the power-law$_{\rm OTHERS}$ 
component, we fitted the spectra and allowed the index to vary freely. 
Then, the index decreased to 1.20, but the abundances 
did not change significantly.
Systematic uncertainty associated with the contaminants on the XIS 
optical blocking filter is less than the statistical errors. 

The derived ICM component in subsection \ref{gal} are including both 
ICM and ISM components.
Considering the point spread function of Suzaku X-ray telescope,
$\sim$ 30 \% of the ICM component in the background region
escaped from the ISM emission in the galaxy region \citep{serlemitsos_07}.
In addition, the ISM emission may extend up to the background region.
Therefore, we added another temperature component in the background 
spectra,
and fitted the spectra again. We got almost the same values for the ISM
component in the galaxy region.
This is because the normalization of the  ICM component in the galaxy 
region
is only 10 \% of that of the ISM, as shown in figure \ref{figure2}.

Because the fits were not formally acceptable, 
we explored using the C-statistic 
giving less biased values than ${\chi^2}$ statistic for 
Poisson-distributed data \citep{humphrey_09}. 
The results of C-statistics fitting hardly improved, ${\chi^2}$/d.o.f.=939/701, 
and the parameters and their errors were almost the same.
The large ${\chi^2}$ mostly comes from the systematic errors of XIS. 
For examples, because the response matrix of XISs around the Si edge has some problems, 
the energy range between 1.82 and 1.84 keV was also ignored
in the spectral fitting \citep{koyama_07}. 
Then, the ${\chi^2}$/d.o.f improved to 906/691.
Furthermore, it is known the discrepancies between the FI and BI
spectra of the background region above 2 keV.
Ignoring this energy range in the background spectra,
the ${\chi^2}$/d.o.f. become 666/578, 
and the parameters were almost the same within a few \%.
The large ${\chi^2}$ are almost due to systematic uncertainties in the background
and calibrations of XIS.

We also investigated the systematic uncertainties in the atomic physics of 
the Fe-L lines as pointed out in \citet{matsushita_00} and \citet{matsushita_07}. 
Convolving the response matrix of XIS, the difference in the Fe-L lines
between APEC \citep{smith_01} and MEKAL \citep{mewe_85} models are typically 10\%.
Therefore, we fitted spectra in the same way as in section \ref{gal} but 
including 10~\% systematic errors within an energy range of 0.75--1.2 keV.
The derived parameters are summarized in table\ref{table2}, and ${\chi^2}$/d.o.f became 859/692. 
The error range in absolute value of each element increased,
and the derived Fe abundance became 0.7--1.8 solar.
However, increases in the error ranges of the abundance ratios
were smaller.
Furthermore, ignoring an energy range of 1.82--1.84 keV,
and above 2 keV in the background spectra, the ${\chi^2}$/d.o.f became 605/569.

\section{Discussion}\label{sec:discuss}

We successfully measured the abundance patterns of O, Ne, Mg, Si, 
and Fe in the ISM of the S0 galaxy NGC~1316 with Suzaku.
Figure \ref{figure4} summarizes the results.
We used the new solar abundance table by \citet{lodders_03}. 
Adopting the result of the 2T model fit,
the abundance pattern in the ISM in NGC~1316 is shown to be close to that of 
the new solar abundance. 
The abundance patterns of the SN II and SN Ia yields are also plotted in figure 
\ref{figure4}.
Here, the SN II yields by \citet{nomoto_06} refer to an 
average over the Salpeter initial mass function of 
stellar masses from 10 to 50 $M_{\odot}$, with a progenitor 
metallicity of $Z=0.02$\@.
The SNe Ia yields  were taken from the W7 model \citet{iwamoto_99}.
The abundance pattern in the ISM of NGC~1316 lies between those 
of SN II and SN Ia, and is consistent with their mixing.
Assuming that the O/Fe ratio of the SN II yield is 3.5 solar 
(\cite{nomoto_06}), then 
$\sim$70\% of Fe in the ISM is synthesized by SNe Ia,
and originates from present SNe Ia and from those trapped in stars.
Since the 2T and 3T models give the Fe abundance of the ISM
of 0.6--1.6 solar, the Fe abundance from SNe Ia would be 0.4--1.1 solar.

The Fe abundance synthesized by present SNe Ia in an early-type 
galaxy is proportional to $M^{\rm Fe}_{\rm SN}\theta_{\rm SN}/\alpha_{*}$ 
(see \cite{matsushita_03} and \cite{nagino2009} for details). 
Here, $M^{\rm Fe}_{\rm SN}$ is the Fe 
mass synthesized in one SN Ia, $\theta_{\rm SN}$ is the SN Ia rate, 
and $\alpha_{*}$ is the stellar mass loss rate. Fe in the hot ISM of the 
galaxy is mainly produced from SNe Ia, since SNe II synthesized far more 
O and Mg than Fe. We used the mass-loss rate from \citet{ciotti_91} and assumed 
the age to be 13 Gyr, which is approximated by $1.5 \times 10^{-11} L_{\rm B}
t^{-1.3}_{15}M_{\odot}\rm{yr}^{-1}$, where $t_{15}$ is the age in units of 
15 Gyr and $L_{\rm B}$ is the B-band luminosity. $M_{\rm Fe}$ produced by one 
SN Ia explosion is likely to be $\sim 0.6M_{\odot}$ \citep{iwamoto_99}. 
Adopting 0.1--0.5 $\rm{SN~Ia}/100\rm{yr}/$ $10^{10}L_{\rm B}$ as the 
optically observed SN Ia rate (e.g., \cite{mannucci_08} and references therein , 
\cite{blanc_04}, \cite{hardin_00}, and \cite{cappellaro_97}), the resultant Fe
abundance is 2.9--14.5 solar, when considering only the SNe Ia contribution. 
Therefore, the expected SNe Ia contribution 
is sufficiently high to supply the Fe in the ISM of this galaxy.

As indicated from the optical 
line indices (\cite{thomas_05};
\cite{bedregal_06}; \cite{bedregal_08}), 
NGC~1316 may contain younger stellar populations. 
In this case, both the stellar mass loss rate and SN Ia rate might be higher 
(\cite{ciotti_91}; \cite{ciotti07}).
The evolution of  SN Ia rate is approximated by power-law model:
$\propto t^{-1.1} L_B$ (\cite{ciotti07}; \cite{greggio05}),
where $t$ is the age of stars.
Then, assuming an age of 3 Gyr, the contribution of SNe Ia to the Fe abundance
decreases by a factor of 0.7.
Since a significant proportion  of stars in NGC~1316 might be older than 3 Gyr, 
the effect of stellar age on the metal-enrichment by present SNe Ia 
would be less than a few tens of a percent.
Then, the difference in the abundance pattern of the ISM of the 
early-type galaxies reflects the difference in the abundance pattern 
of the stars in the entire galaxy.

Since a longer star-formation time-scale yields more SNe Ia products in 
stars, the differences in the $\alpha$/Fe ratio in stars can constrain 
the star-formation histories.
The abundance pattern of O, Ne, Mg, Si, and Fe in the ISM of NGC~1316
is close to that of the elliptical galaxies, have been obtained 
by Chandra, XMM, and Suzaku (e.g.\cite{humphrey_06}, \cite{werner_06}, and \cite{matsushita_07}).
We show the abundance patterns of 
NGC~4382, NGC~1332, NGC~720, NGC~1399, NGC~1404, NGC~4636, and NGC~5044 researched previously
(\cite{nagino2009}, \cite{humphrey_04}, \cite{tawara_08}, \cite{matsushita_07}, 
\cite{hayashi_09}, \cite{buote_03}, and \cite{komiyama_09}) in figure \ref{figure4}.
Considering the expected high contribution from the present SNe Ia, 
the observed 
solar abundance pattern in the ISM of NGC~1316 indicates an overabundance 
of $[\alpha/{\rm Fe}]$ in
stars in the entire galaxy as in these elliptical galaxies, although
optical observations indicate that stars in the central 1/10 $r_e$ region 
of NGC~1316 contain more SN Ia products than those in the giant elliptical 
galaxies \citep{thomas_05}.
Suzaku observed the ISM of the entire galaxy and revealed the 
abundance pattern in stars over the entire galaxy.

The abundance pattern of the S0 galaxy, NGC~4382 was also investigated with Suzaku
(\cite{nagino2009}). 
NGC~4382 is a normal S0 galaxy located on the outskirts of the Virgo cluster.
Figure \ref{figure4} also shows that the  O/Fe ratio of the ISM in NGC~4382 
is smaller, by a factor of two, as compared to that of NGC~1316.
The different abundance patterns of these two S0 galaxies 
reflect a higher amount of SNe Ia products in the ISM of NGC~4382.
Measurements of optical line indices indicate that 
stars in the most central region ($<1/8 r_e$) of these two galaxies have similar
$[\alpha/{\rm Fe}]$ ratios (\cite{thomas_05}; \cite{McD2006}).
As shown above, the effect of stellar age on the Fe abundance of the 
ISM from the present SNe Ia might be small, but 
stars in the entire NGC~4382 may contain more SNe Ia products
than those in the entire NGC~1316.
These results suggest differences in the metal enrichment histories and hence
in the formation and evolution histories of the
ISM in the merger remnant, NGC~1316, and the normal S0 galaxy, NGC~4382.
Optical observations reveal that the $\alpha/{\rm Fe}$ 
ratio of the stars in the central regions 
of elliptical and S0 
galaxies depends on the system mass (e.g., \cite{thomas_05}; \cite{bedregal_08}).
The temperature of the ISM of NGC~4382 is about 0.3 keV 
(\cite{nagino2009}) while and those of NGC~1316 and the ellipticals
are more than 0.5 keV, as shown in figure \ref{figure4}.
Since the temperature of the ISM reflects the system mass,
the differences in the abundance patterns in the ISM of the two galaxies
may also reflect the dependence of the $\alpha/{\rm Fe}$ ratio of
stars in the entire galaxy on the system mass. 
In addition, the abundance pattern of another S0 galaxy, NGC~1332 was 
searched with XMM and Chandra \citep{humphrey_04}. 
Although the temperature of the ISM of NGC~1332 
is $\gtrsim$ 0.5 keV, the $\alpha/{\rm Fe}$ has large error to compare with 
other S0 galaxies, as shown in figure \ref{figure4}.
Further studies to obtain, 
more samples of the elliptical and S0 galaxies with different ISM 
temperatures would enable us to understand the formation process of 
the early-type galaxies in more detail.

\bigskip
We thank the referee for providing valuable comments.
We gratefully acknowledge all members of the Suzaku hardware and software 
teams and the Science Working Group. 
SK is supported by JSPS Research Fellowship for Young Scientists.

\end{document}